# Voltage control of magnetic anisotropy in epitaxial Ru/Co$_2$FeAl/MgO heterostructures

Zhenchao Wen,[1] Hiroaki Sukegawa,[1] Takeshi Seki,[2,3] Takahide Kubota,[2,3] Koki Takanashi,[2,3] and Seiji Mitani[*1,4]

[1]*National Institute for Materials Science (NIMS), Tsukuba 305-0047, Japan*
[2]*Institute for Materials Research (IMR), Tohoku University, Sendai 980-8577, Japan*
[3]*Center for Spintronics Research Network (CSRN), Tohoku University, Sendai 980-8577, Japan*
[4]*Gradate School of Pure and Applied Sciences, University of Tsukuba, 305-8577, Japan*

**Abstract:**

Voltage control of magnetic anisotropy (VCMA) in magnetic heterostructures is a key technology for achieving energy-efficiency electronic devices with ultralow power consumption. Here, we report the first demonstration of the VCMA effect in novel epitaxial Ru/Co$_2$FeAl(CFA)/MgO heterostructures with interfacial perpendicular magnetic anisotropy (PMA). Perpendicularly magnetized tunnel junctions with the structure of Ru/CFA/MgO were fabricated and exhibited an effective voltage control on switching fields for the CFA free layer. A large VCMA coefficient of 108 (139) fJ/Vm for the CFA film was achieved at room temperature (4 K). The interfacial stability in the heterostructure was confirmed by repeating measurements. Temperature dependences of both the interfacial PMA and the VCMA effect were also investigated. It is found that the temperature dependences follow power laws of the saturation magnetization with an exponent of ~2. The significant VCMA effect observed in this work indicates that the Ru/CFA/MgO heterostructure could be one of the promising candidates for spintronic devices with voltage control.

*Email: Mitani.Seiji@nims.go.jp

Achieving a sufficient voltage control of magnetic anisotropy (VCMA) in magnetic heterostructures is of particular importance for realizing energy-efficiency electronic devices with ultralow power consumption, such as voltage-torque magnetoresistive random access memories (MRAMs)[1,2] and low power logic LSIs.[3,4] By using voltage-controlled magnetization switching, the power consumption for manipulating a bit cell is proposed to be ~1 fJ, which is two orders of magnitude lower than ~0.1 pJ in the spin transfer torque (STT) technology.[5] Since magnetic tunnel junctions (MTJs) are key building blocks for high-performance spintronic devices,[6] the voltage control of magnetization switching in MTJs has greatly promoted the development of the VCMA technology to practical applications.[7–17] Recently, a ultralow power consumption of ~6 fJ is achieved in state-of-the-art CoFeB/MgO MTJs with voltage-controlled magnetization switching.[13,14] Regarding the origin of the VCMA effect, several physical mechanisms have been proposed, such as the modulation of spin-orbit interactions by charge accumulation and depletion,[18] Rashba effect,[19,20] voltage-induced redox reaction,[21] and electro-migration.[22] Designing and exploring materials for realizing a large magnitude of VCMA effect are still in the ascendant. To date, well-engineered CoFeB,[8–14] CoFe,[7,15] FeB,[16] Fe,[17] and FePt[23,24] ultrathin layers combining with MgO insulators have been extensively studied; however, the voltage effect on the change of magnetic anisotropy is still not large enough in the material systems and the bit-error rate is also a challenge for mass products.[25] In addition, a large VCMA effect with changing magnetic anisotropy of ~5000 fJ/Vm was observed in Co/GdOx bilayers by voltage-driven $O^{2-}$ migration;[21] whereas, the speed of ions migration process is very slow and the endurance and scalability are still under discussion. Therefore, advanced materials are extremely required for further development of voltage-controlled spintronic devices.

Ru/Co$_2$FeAl(CFA)/MgO is a novel magnetic heterostructure exhibiting a large perpendicular magnetic anisotropy (PMA) induced by the CFA/MgO interface.[26–28] A high out-of-plane tunnel magnetoresistance (TMR) ratio of more than 130% was obtained in fully stacked CFA/MgO MTJs on the 4-fold-symetry Ru buffer at room temperature (RT).[28] In addition, it was reported that the CFA alloy film can possess a low intrinsic damping constant,[29] which is greatly advantageous for decreasing bit-error rate for voltage-controlled spintronic devices.[25] Recently, giant VCMA effect as high as 1000 fJ/Vm in the CFA/MgO structure was theoretically proposed by first principles calculations.[30] However, from an experimental point of view, complicated interface reactions and/or inter-diffusions could occur in practical ultrathin heterostructures, such as boron diffusion in Ta/CoFeB/MgO structures,[31,32] which is particularly significant for both the interfacial PMA and the VCMA effect. Therefore, experimental examination of the VCMA effect in the Ru/CFA/MgO heterostructures is desired for practical spintronic applications, as well as for promoting theoretical investigations on the origin of the VCMA effect.

In this work, we experimentally demonstrate the VCMA effect in Ru/CFA/MgO MTJs. A large coercivity change induced by voltages was observed for the CFA layer in perpendicularly magnetized MTJs (p-MTJs) with the structure of CFA/MgO/CoFeB/Ta/[TbCo/Co]$_{25}$ where the CFA was a free layer and the CoFeB/Ta/[TbCo/Co]$_{25}$ composited layer acted as a reference layer. Quantitative analysis of the VCMA effect was performed by utilizing a magnetoresistance measurement in Ru/CFA/MgO/CoFeB MTJs with an orthogonally magnetic configuration.[33] A large VCMA coefficient of 108 (139) fJ/Vm was achieved for the CFA film in the epitaxial Ru/CFA/MgO structure at RT (4 K). Interfacial stability in the structure was demonstrated by repeating measurements of the VCMA effect. Also, the dependences of both interfacial PMA and the VCMA effect on temperature were investigated. It is found that both of them follow power laws of saturation magnetization.

Figure 1a illustrates the whole stack structure for the designed p-MTJs. The deposition condition for each layer is also indicated and the positive direction of applied voltage is defined as the electric current flows from top to bottom layers. The composite upper layers with the structure of CoFeB/Ta/[TbCo/Co]$_{25}$ are as a reference layer since the TbCo alloy was reported to have a large coercivity,[34] while the 1-nm-thick CFA is a free layer. The utilization of CoFeB/Ta bilayers in the reference layer is to maintain a high TMR ratio by coherent tunneling through the MgO barrier. The tunneling resistance ($R$) as a function of out-of-plane magnetic field ($H$) for a patterned p-MTJ with 100-nm-nanoscale in diameter is shown in Fig. 1b. A direct-current (dc) voltage of 1 mV was applied to measure the resistance here. A TMR ratio of 65% was achieved with a resistance-area product ($RA$) of ~175 $\Omega \cdot \mu m^2$. The TMR ratio is defined as $100 \times (R_{AP} - R_P) / R_P$, where $R_{AP}$ ($R_P$) is the resistance of the antiparallel, AP (parallel, P) magnetization state. A sharp resistance change observed in the middle of the full $R-H$ loops indicates that a perfect perpendicular magnetization and coherent switching for the CFA free layer were achieved in the heterostructure. Figure 1c shows the minor loops of normalized TMR as a function of magnetic field at the bias voltages of −800 mV, 1 mV, and 800 mV, respectively. A large change in the switching field ($H_s$) is observed in the CFA layer when voltages were applied. For instance, the $H_s$ from P to AP state is −780 Oe at 800 mV while it becomes −900 Oe at −800 mV. The $\Delta H_s$ of 120 Oe is modulated by the voltages, which is much larger than that in CoFeB/MgO p-MTJs.[8] The detailed voltage dependence of switching fields for the CFA layer from P (AP) to AP (P) magnetization states is shown in Fig. 1d. The coercivities of the CFA layer at each voltage can be derived by $(H_{s(AP-to-P)} - H_{s(P-to-AP)})/2$. It is found that the coercivity decreases with increasing voltage from negative to positive bias direction, which indicates the mechanism of the VCMA effect in the p-MTJ could be mainly from charge accumulation/depletion.[18]

The charge accumulation/depletion modulates the electronic occupancy of 3*d* orbitals in the band structure as well as spin-orbit interactions at the CFA/MgO interface, resulting in the change in effective magnetic anisotropy of the CFA layer.

In order to quantitatively evaluate the VCMA effect in the Ru/CFA/MgO heterostructures, we use a method of in-plane magnetoresistance measurement by utilizing an orthogonally magnetized MTJ.[33] Figure 2a shows the illustration of the MTJ structure in which the CFA layer is perpendicularly magnetized while the CoFeB layer has in-plane magnetic anisotropy. The external magnetic field was applied along the in-plane direction and *R*−*H* loops were measured under different voltages. The VCMA effect was estimated by normalizing the measured curves using the minimum (*H* = 10 kOe) and maximum (*H* = 0 Oe) resistances. Figure 2b shows normalized TMR curves at the voltages of 10, ±500, and ±1100 mV in the positive magnetic field region. A clear dependence of TMR curves on the bias voltages was observed. Since the influence of the bias voltages on the magnitude of TMR ratio is excluded by the normalizing process, the variation in the TMR curves at different voltages indicates a remarkable VCMA effect on the CFA layer.

The PMA of the CFA films was calculated using the measured *R*−*H* curves. The conductance of the MTJ is related to the relative angle $\theta$ between the magnetizations of CFA and CoFeB layers, which can be expressed by the equation below,

$$G(H) = G_{90°} + (G_{0°} - G_{90°}) \cos\theta. \qquad (1)$$

Here $G_{90°}$ ($G_{0°}$) = $1/R_{90°}$ ($1/R_{0°}$) is the conductance of the MTJ in orthogonal (parallel) magnetization configuration. Because the magnetization of CoFeB reference layer is fixed in the film plane, the in-plane component of the magnetization of the CFA layer can be obtained by,

$$M_{in-plane}(H) = M_S \cos\theta = M_S \frac{G(H) - G_{90°}}{G_{0°} - G_{90°}} \qquad (2)$$

where $M_S$ is the saturation magnetization of the CFA film. Furthermore, the PMA energy density $K_u$ can be calculated as,

$$K_u = \mu_0 M_S \int_0^1 H dM_{\text{norm.}} \tag{3}$$

where $\mu_0$ is the permeability of free space and $M_{\text{norm.}}$ is the $M_{\text{in-plane}}$ normalized by $M_s$. A typical normalized in-plane component of the magnetization is shown in Fig. 2c. The shaded area indicates the magnitude of PMA energy density. Figure 2d shows the PMA density per unit area, $K_u t$, as a function of applied electric field for the CFA film. At the negative (positive) bias direction, an increase (decrease) in the $K_u t$ with increasing the magnitude of electric field was observed. A clear deviation of linear tendency of the $K_u t$–$E$ curve between positive and negative bias directions was shown. The similar behavior was also observed in the CoFeB/MgO[8] and Fe/MgO[17] heterostructures. Linear fits were performed independently for the positive and negative bias regions. The slope of 108 (50) fJ/Vm was achieved in the negative (positive) bias direction, which is defined as the coefficient $\xi$ of the VCMA effect in the CFA film. The values are much larger than 33 fJ/Vm for the CoFeB film in the structure of Ta/CoFeB/MgO.[10] Also, we were aware of opposite signs of VCMA effect, i.e., the sign of $\xi$, observed between polycrystalline Ru/CoFeB/MgO and Ta/CoFeB/MgO structures.[10] Interestingly, the sign of the VCMA effect in this work is opposite to that of the polycrystalline Ru/CoFeB/MgO and is the same as that of the Ta/CoFeB/MgO. The sign of the VCMA effect is expected to be sensitive to the electronic structures near the interfaces of a ferromagnetic (FM) layer. Especially, the electronic structures of nonmagnetic (NM) layers adjacent to the FM layer can determine the signs of the VCMA effect, which had been demonstrated by first principles calculations in Pt/Fe/Pt and Pd/Fe/Pd structures.[35,36] Therefore, the difference in the ferromagnetic material (CFA and CoFeB) and/or crystal structure (e.g., polycrystalline hcp Ru and single crystalline

Ru(02$\bar{2}$3) buffer layers) may be responsible for the discrepancy of the signs. However, the buffer layer dependence of the sign of the VCMA effect, as well as the deviation of the VCMA magnitudes between negative and positive bias regions,[17] are still under discussions.

Furthermore, we investigated the chemical stability of the interfaces in the Ru/CFA/MgO heterostructure by repeatedly measuring the VCMA effect. Figure 3 shows typical VCMA measurements by applying bias voltages of ±1 V back and forth for 300 times. The time interval between two points is one second. The PMA energy density of $K_u t$ is calculated at each bias voltage for every time. The inset is a magnified figure for 50-times measurements. The PMA value is well maintained without any decay in the count scale, indicating the crystal structures and interfaces are chemically stable in the Ru/CFA/MgO heterostructure. This result also implies that the VCMA effect in the system does not originate from the ionic migration or defects in the materials.

Next, we study the temperature dependence of the VCMA effect. Figure 4a shows the applied electric field dependence of $K_u t$ at different temperatures. The $K_u t$ is calculated by the equation (3) in which the $M_s$ at each temperature is estimated by Bloch's law with the Curie temperature $T_c$ = 1170 K and $M_s$ (300 K) = 1050 emu/cm$^3$,[26,37]

$$M_s(T) = M_s(0\text{K})[1 - (\frac{T}{T_c})^{3/2}]. \qquad (4)$$

With decreasing temperature, a clear shift of $K_u t - E$ curves is shown, indicating an increase in the magnetic anisotropy for the CFA film at low temperatures. The discrepancy in the linear slopes of the $K_u t - E$ curves between positive and negative bias regions was still observed in the entire temperature range. Figure 4b shows the interfacial magnetic anisotropy $K_u t$ at zero applied electric field as a function of temperature. The solid line indicates a fitted curve by a power law of $M_s(T)$,

$$K_u t(T) = K_u t(0\text{K})(\frac{M_s(T)}{M_s(0\text{K})})^\gamma. \qquad (5)$$

It is found that the temperature dependence of $K_u t$ is well fitted by the power law with an exponent of $\gamma = 2.00 \pm 0.08$. The result is corresponding with the theoretical prediction of a $M_s^2(T)$ dependence for the magnetic anisotropy especially in the material systems with PMA contributed by high spin-orbit interactions.[38,39] In general, the temperature dependence of uniaxial anisotropy follows Callen-Callen's $M_s^3(T)$ law in a simple ferromagnetic system.[40] The $M_s^2(T)$ power law indicates that the Ru/CFA/MgO system processes a strong spin-orbit interaction where localized moment or two-ion anisotropic exchange makes a main contribution to the temperature dependence of magnetic anisotropy.[38,39] A similar $M_s^2(T)$ dependence was also observed in MgO/CoFeB/Ta heterostructures with interfacial PMA.[41] Furthermore, we investigate the VCMA coefficient $\xi$ as a function of temperature, as shown in Fig.4c. The $\xi_{posi.}$ ($\xi_{nega.}$) values are derived by linearly fitting the $K_u t - E$ curves in the positive (negative) regions. A maximum $\xi$ of 139 fJ/Vm is achieved at 4 K for the negative region. The VCMA coefficient $\xi_{nega.}$ decreases with increasing temperature while the $\xi_{posi.}$ is nearly independent of the temperature. We fit the temperature dependence of $\xi_{nega.}$ using a power law of $M_s(T)$,

$$\xi(T) = \xi(0K)\left(\frac{M_S(T)}{M_S(0K)}\right)^{\gamma'}. \tag{6}$$

The solid line in the Fig. 4c shows a well fitted curve by the power law of $M_s(T)$ with an exponent $\gamma = 1.89 \pm 0.29$, which is near to the $M_s^2(T)$ as the temperature dependence of interfacial magnetic anisotropy. The exponent is smaller than 2.83 in the MgO/CoFeB/Ta system,[41] which could be attributed to the different spin-orbit coupling strength in Fe-O hybridization and/or discrepant electronic band structures at the FM/NM interfaces. Figure 4d shows the magnetic anisotropy dependence of the VCMA effect. The $\xi_{nega.}$ obtained in the negative bias region increases with increasing $K_u t$, whereas the $\xi_{posi.}$ for the positive bias region shows a weak magnetic anisotropy dependence. The deviation of the dependences of $\xi$ on $T$ and $K_\mathrm{u}t$ between positive and negative bias directions could be attributed to the electronic band structures at the Ru/CFA/MgO interfaces. Theoretical calculations will be strongly required for further understanding of the mechanisms to explain the observed temperature and magnetic anisotropy dependences of the VCMA effect.

In summary, the VCMA effect in the Ru/CFA/MgO heterostructures was experimentally demonstrated. Perpendicularly magnetized MTJs with the structure of CFA/MgO/CoFeB/Ta/[TbCo/Co]$_{25}$ were fabricated, which exhibit effectively voltage-controlled switching fields for the CFA free layer. Quantitative analysis of the VCMA effect reveals a large VCMA coefficient of 108 (139) fJ/Vm for the CFA film in the epitaxial Ru/CFA/MgO structure at RT (4 K). The chemical stability of the heterostructure is examined by repeating VCMA measurements. It is also investigated that temperature dependences of both the interfacial PMA and the VCMA effect follow power laws of the saturation magnetization. This study indicates the Ru/CFA/MgO heterostructure could be a promising candidate for spintronic applications with voltage-controlled magnetism.

## Methods

All the multilayers were deposited by magnetron sputtering in an ultrahigh vacuum system with a base pressure of $3 \times 10^{-7}$ Pa. The p-MTJ multilayer stack has a structure of MgO(001)-substrate//Ru(40)/CFA(1)/MgO(1.3)/CoFeB(1.5)/Ta(0.45)/[TbCo(0.3)/Co(0.3)]$_{25}$/Ta(5)/Ru(10) (unit: nm). The Ru layer was post-annealed at 400 °C for 30 min after deposition in order to form an ideal $(02\bar{2}3)$ epitaxial growth for achieving high-performance CFA/MgO bilayers. The bottom stack of Ru/CFA/MgO/CoFeB/Ta was annealed at 325 °C for 1h before depositing TbCo/Co multilayers for the purpose of obtaining both high TMR and large PMA simultaneously. The TbCo/Co multilayers were deposited at RT by a co-sputtering method from Co and Tb targets. Furthermore, an orthogonally magnetized MTJ stack with the structure of MgO(001)-substrate//Ru/CFA(1)/MgO(2.2)/CoFeB(20)/Ta/Ru (unit: nm) was deposited for quantitative analysis of VCMA effect on the CFA layer. The condition of the deposition is identical to that for p-MTJs. All the MTJ stacks were patterned into junctions with electric contacts by electron beam lithography and conventional UV lithography combining with Ar ion milling and lift-off technique. The magneto-electrical transport measurements were performed by a dc four-probe method at RT and low temperatures in a physical property measurement system (PPMS).

## Acknowledgements

This work was partly supported by the ImPACT program of the Council for Science, Technology and Innovation (Cabinet Office, Government of Japan) and JSPS KAKENHI Grant Number 16H06332. A part of this work was performed under the Inter-university Cooperative Research Program of Institute for Materials Research, Tohoku University (16K0072).

## Author Contributions

S.M. and K.T. supervised the study. Z.W. and S.M. designed the experiments. Z.W. and H.S. deposited and micro-fabricated the samples. Z.W., T.S. and T.K. carried out the measurements and analyzed the data. Z.W. and S.M. wrote the manuscript with reviewing by H.S., T.S., T.K. and K.T.. All authors discussed the results and the manuscript.

## Additional Information

Competing financial interests: The authors declare no competing financial interests.

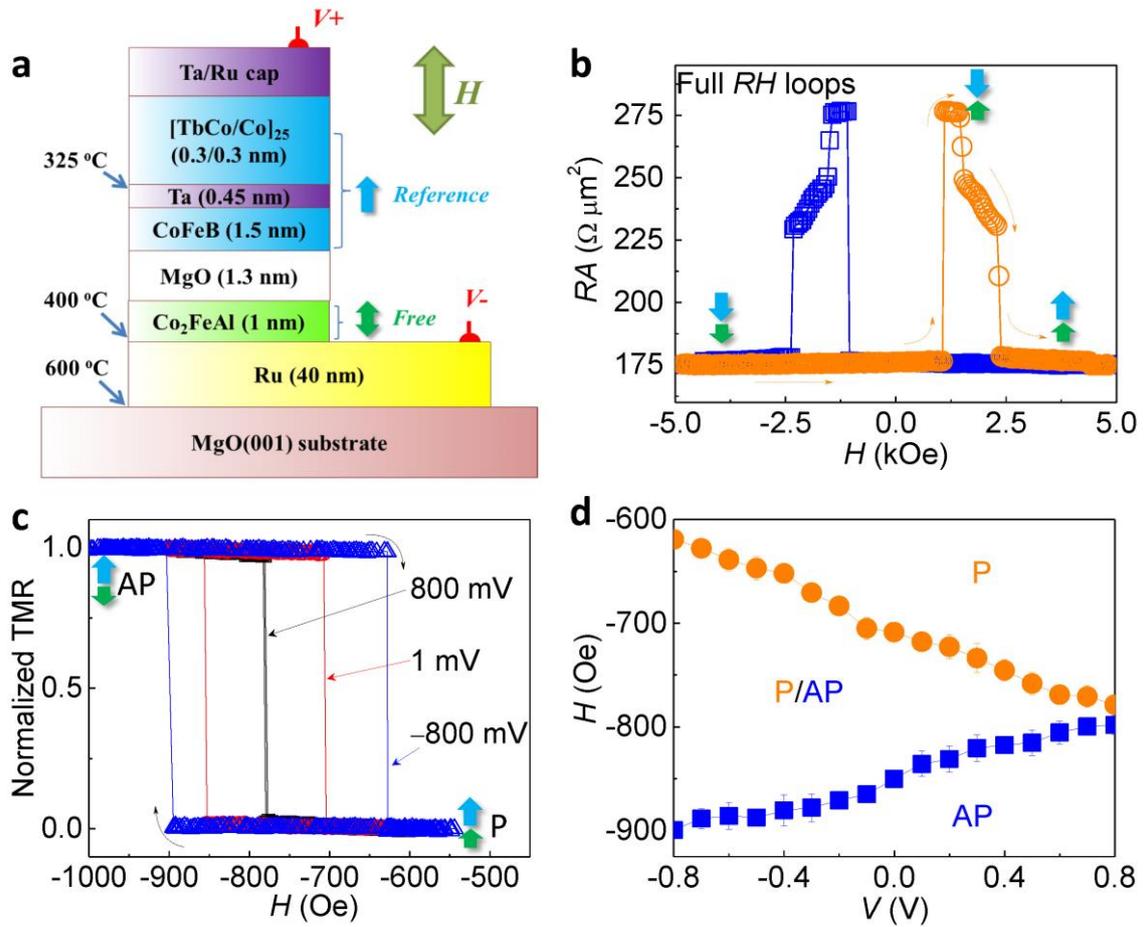

**Figure 1. Demonstration of VCMA effect in p-MTJs.** (**a**) Schematic illustration of the whole stack structure for p-MTJs. (**b**) Full *R–H* loops of a nanoscaled p-MTJ with applying an electric voltage of 1 mV. (**c**) Minor *R–H* loops for the p-MTJ under the voltages of −800, 1, and 800 mV. (**d**) Magnetic phase diagram for the CFA layer from P (AP) to AP (P) magnetic states dependent on voltage and magnetic field. All the measurements were performed at RT.

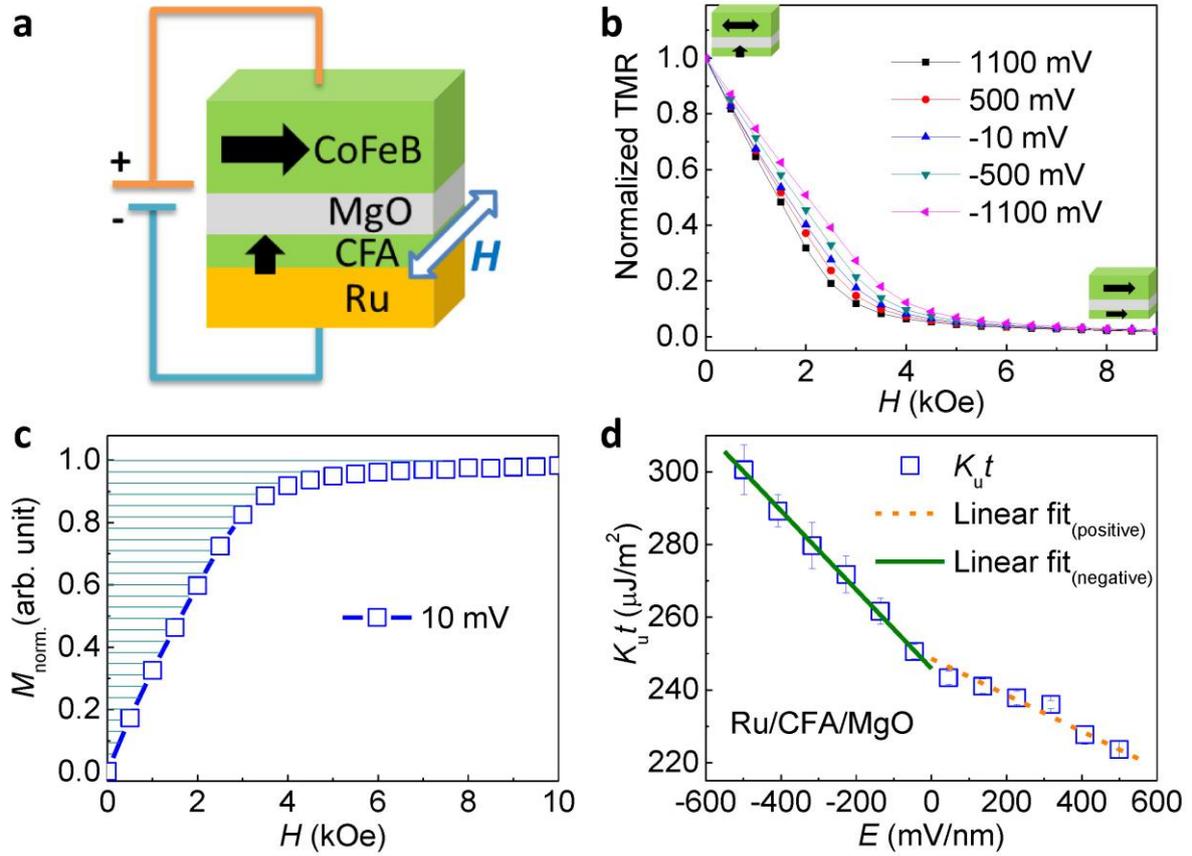

**Figure 2. Quantitative estimation of VCMA effect in the Ru/CFA/MgO heterostructures.** (**a**) Schematic illustration of orthogonally magnetized MTJs with the structure of Ru/CFA/MgO/CoFeB. (**b**) Normalized TMR curves obtained at different applied voltages under in-plane magnetic field (only the positive field region is shown here). (**c**) Typical normalized in-plane component of the CFA magnetization. The PMA energy density can be estimated by the shaded area. (**d**) Electric field dependence of the magnetic anisotropy $K_u t$ for the CFA film.

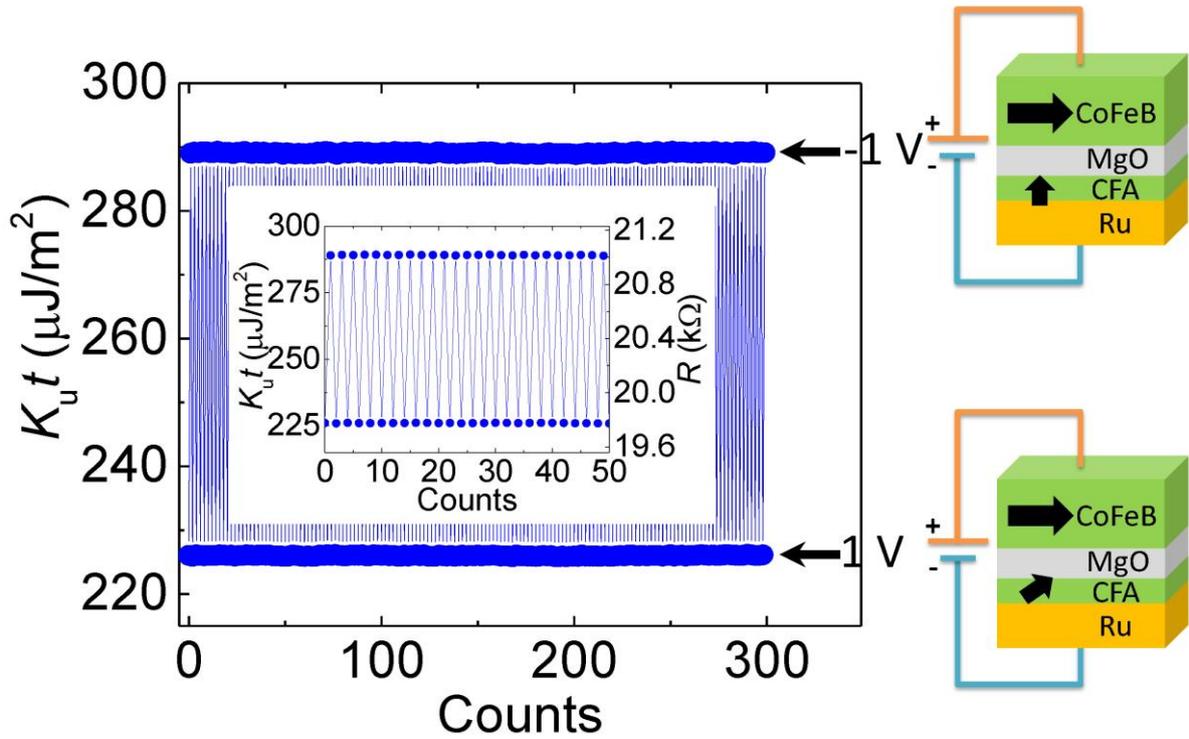

**Figure 3. Repeating measurements of the VCMA effect.** Bias voltages of ±1 V were alternately applied on the MTJs for 300 times. The illustrations indicate the magnetic configuration in the MTJ when a voltage is applied. The inset is the magnification for 50-times measurements.

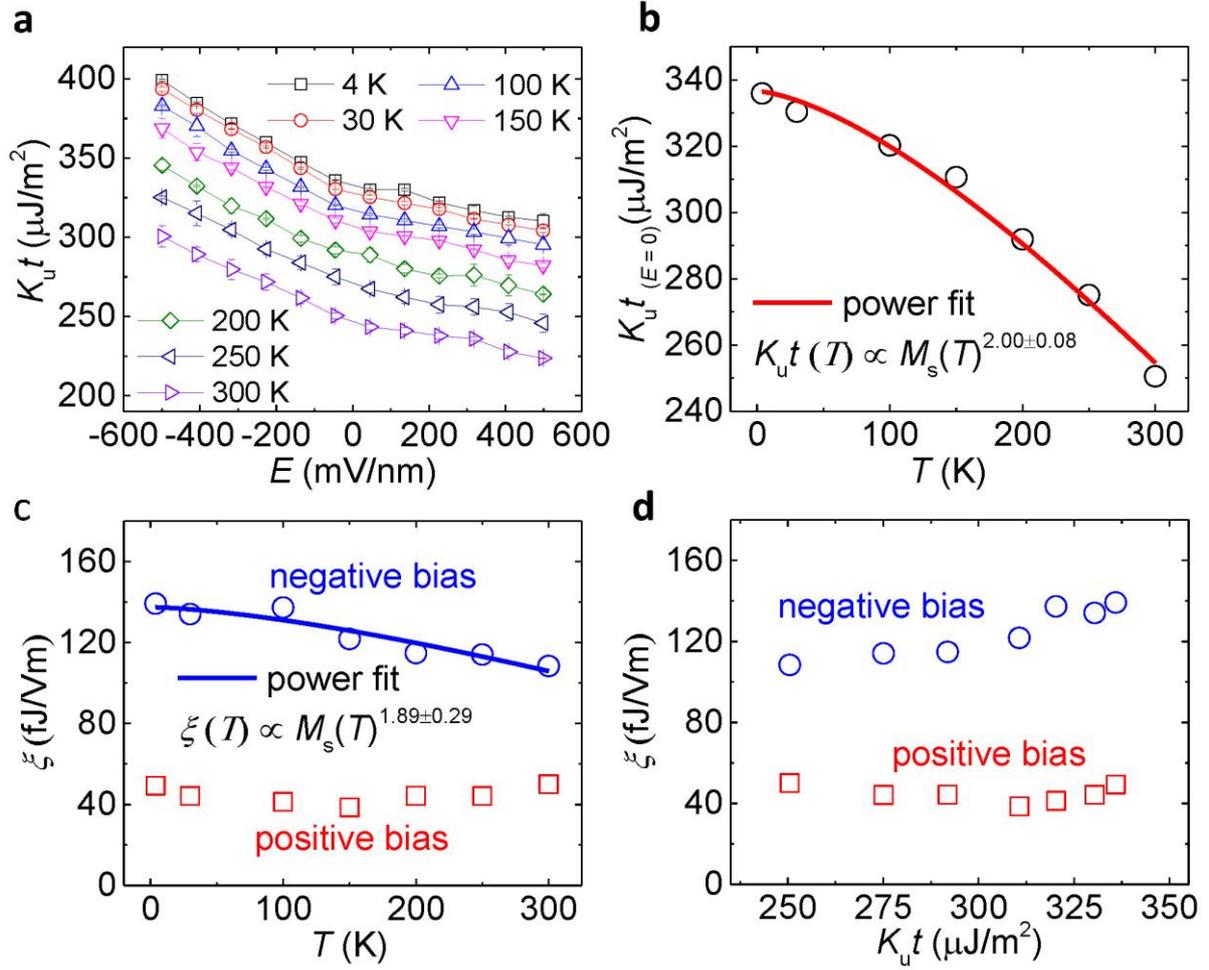

**Figure 4. Temperature and magnetic anisotropy dependence of VCMA effect in the Ru/CFA/MgO heterostructures.** (**a**) The dependence of $K_u t$ on applied electric field at different temperatures. (**b**) Temperature dependence of interfacial magnetic anisotropy at zero voltage. (**c, d**) The VCMA coefficient $\xi$ as a function of (**c**) temperature and (**d**) magnetic anisotropy for the CFA film in the negative and positive bias regions. The solid lines in (**b**) and (**c**) are fitted curves by power laws of $M_s(T)$ with exponents (**b**) $\gamma = 2.00 \pm 0.08$ and (**c**) $\gamma = 1.89 \pm 0.29$, respectively.